\def \imat {{\rm i}}

\documentstyle[twocolumn,floats,epsf,ifthen,aps,prl]{revtex} 

\begin{document} 

%\preprint{draft} 
 
%\draft 

\title{Noisy quantum measurement of solid-state qubits:
 Bayesian approach} 

\author{Alexander N. Korotkov} 
\address{ 
Department of Electrical Engineering, University of California, 
Riverside, CA 92521-0204. 
} 
\date{\today} 
 
\maketitle 
 
\begin{abstract} 
We discuss a recently developed formalism which describes 
the quantum evolution of a solid-state 
qubit due to its continuous measurement. In contrast to the conventional
ensemble-averaged formalism, it takes into account the measurement 
record and therefore is able to consider individual realizations of the 
measurement process. The formalism provides testable experimental
predictions and can be used for the analysis of a quantum feedback 
control of solid-state qubits. We also discuss  generalization 
of the Bayesian formalism to the continuous measurement of 
entangled qubits. 
\end{abstract} 
%\pacs{PACS numbers: }
%\pacs{73.23.-b; 03.65.Ta; 03.67.Lx}
 
\narrowtext 
 
%\vspace{1ex} 
\vspace{0.6cm}

\section{Introduction}

        Bayesian approach to the problem of continuous quantum 
measurement is a relatively new subject in solid-state mesoscopics, 
even though this approach has a long history \cite{Davies,Kraus}
as a general quantum framework 
and is rather well developed, for example, in quantum optics 
\cite{Wiseman-93} 
(for more references, see Ref.\ \cite{Kor-rev}) 
        The main problem considered in this paper is a very simple 
question: {\it what is the 
evolution of a quantum two-level system (qubit) during the process of
its measurement by a solid-state detector } (Fig.\ 1)? 
In spite of the question simplicity, the answer is not that trivial.

The textbook ``orthodox'' quantum mechanics \cite{Neumann} says that 
the measurement should instantly collapse the qubit state, so that 
after the measurement the qubit state is either $|1\rangle$ or $|2\rangle$,
depending on the measurement outcome.  [The  measurement  basis  is 
obviously 
defined by the detector; in particular, it is a charge basis for the
examples of Figs.\ 1(a) and 1(b).] Such answer is sufficient for typical 
optical experiments when the measurement  is instantaneous (a scintillator
flash or a photocounter click). 
However, for typical solid-state setups (as well as for some more advanced 
setups in quantum optics \cite{Wiseman-93}) 
the instantaneous collapse is not a sufficient answer. 
In particular, in the examples of Fig.\ 1 typically  the detector is 
weakly coupled to the qubit, so the measurement process can take
a significant time and therefore the collapse should be considered
as a continuous process. The notion of a continuous evolution 
due to measurement is well accepted in the solid-state community 
and is usually considered within the framework of the Leggett's 
formalism \cite{Caldeira,Zurek}. This formalism gives the decoherence-based 
answer to the question posed above. It says that the nondiagonal matrix 
elements of the qubit density matrix (obtained by tracing over the detector
degrees of freedom) gradually decay to zero, while the diagonal matrix
elements do not evolve (assuming that the qubit does not oscillate 
by itself, $H=0$, where $H$ describes the tunneling between $|1\rangle$
and $|2\rangle$). 
So, after the completed measurement we have an incoherent mixture of 
the states $|1\rangle$ and $|2\rangle$. 

        Let us notice that these two answers to our question obviously 
{\it contradict} each other and the ``orthodox'' answer cannot be obtained 
as some limiting case of the decoherence answer (since decoherence does
not lead to localization into one definite state). The resolution of 
the apparent contradiction is simple: two approaches consider different
objects. The decoherence approach describes the {\it average evolution 
of the ensemble} of qubits, while the ``orthodox'' quantum mechanics 
is designed to treat a {\it single quantum system}. This difference also
explains the inability of the decoherence formalism to take 
the measurement outcome into account. 

\begin{figure}
\centerline{
\epsfxsize=3.2in 
\hspace{-0.1cm}
\epsfbox{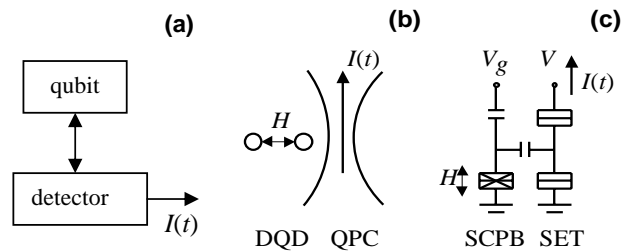}}  
\vspace{0.3cm} 
\caption{(a) General schematic of a continuously measured solid-state 
        qubit and two particular realizations of the setup: 
        (b) a qubit made of double quantum dot (DQD) measured by a 
        quantum point contact (QPC) and (c) a qubit based on 
        single-Cooper-pair box (SCPB) measured by a single-electron 
        transistor (SET).
        }
\label{setup} \end{figure} 

        Obviously, it is desirable to have a formalism which would combine 
advantages of the two approaches and describe the 
{\it continuous measurement of a single qubit}. Then the ``orthodox''
result would be a limiting case for very fast (and ``strong'') measurement, 
while the decoherence result could be obtained by an ensemble averaging. 
The Bayesian formalism \cite{Kor-rev,Kor-99} which is the subject
of this paper has been developed exactly for that purpose 
(some extensions of the Bayesian formalism will be discussed later). 
Notice that the formalism has
been also reproduced in a somewhat different language (using the 
terminology of quantum trajectories, quantum jumps, 
and quantum state diffusion) by another group \cite{Goan}. 
        It is important to stress that the Bayesian approach is not 
a phenomenological formalism which just correctly describes two 
previously known cases. It claims the description of a real and
experimentally verifiable evolution of a single qubit in a process
of measurement. 

        Simply speaking, the Bayesian formalism gives the following answer 
to the question posed above (for $H=0$). During the 
measurement process the diagonal matrix elements of the qubit density 
matrix evolve according to the classical Bayes 
formula \cite{Bayes,Borel} which takes into account the noisy detector 
output [$I(t)$ in Fig.\ 1] and describes a gradual qubit localization into
one of the states $|1\rangle$ or $|2\rangle$, depending on $I(t)$. 
The evolution of nondiagonal matrix elements can be easily calculated 
using somewhat surprising result that a good (ideal) detector preserves the 
purity of the qubit state, so that the decoherence is actually just a 
consequence of averaging over different detector outcomes $I(t)$ for 
different
members of the ensemble. (Nonideal detectors also produce some amount of 
qubit decoherence, which is calculated within the formalism.) 

Notice that in the case 
of an ideal detector, our result can actually be considered as a simple 
consequence of the so-called Quantum Bayes Theorem (we borrow this name from 
the book on quantum noise by Gardiner \cite{Gardiner}, even though 
it is not a theorem in a mathematical sense). However, 
the application of this ``theorem'' is not always straightforward, 
so instead of applying it as an ansatz, we derive the Bayesian formalism 
for particular measurement setups, starting from the textbook quantum 
mechanics. 

	It is difficult to avoid philosophical questions discussing 
a problem related to quantum measurements. In brief, 
philosophy of the Bayesian approach is exactly the philosophy of 
the ``orthodox'' quantum mechanics. A minor technical difference is that 
instead of assuming instantaneous information on 
measurement result corresponding 
to instantaneous
``orthodox'' collapse, we consider a more realistic case of continuous
information flow.

        Finally, let us mention that the problem of solid-state qubit 
evolution due to continuous measurement was recently a subject of 
theoretical study by many groups (see, e.g.\ 
\cite{Gurvitz-97,Gurvitz-98,Stodolsky,Averin,Makhlin,Hackenbroich}). 
However, most of them assumed ensemble averaging and so obtained results
different from the Bayesian results (except the Australian group 
\cite{Goan,Goan-2,Goan-3} which also studies single realizations of 
the measurement process).

\section{Simple model} 

\begin{figure} 
\centerline{
\epsfxsize=2.5in 
\epsfbox{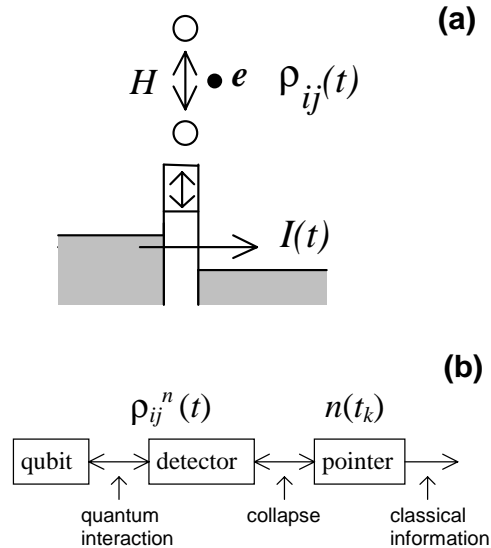}
} 
\vspace{0.3cm}
\caption{(a) Tunnel junction as a detector of the double-quantum-dot qubit. 
The electron location in the DQD affects the detector barrier height. 
The noisy current $I(t)$ (detector output) reflects the evolution of 
the qubit density matrix $\rho_{ij}(t)$.
(b) Idea of the Bayesian formalism derivation via Bloch equations. The
number $n$ of electrons passed through the detector is periodocally
collapsed (forced to choose a definite value) at moments $t_k$.
 }
\label{Schematic}\end{figure}

        Even though Bayesian approach is applicable to a broad range 
of measurement setups, let us start with a particularly simple setup
[Fig.\ 1(b)] consisting of a double quantum dot occupied by a single 
electron, the position of which is measured by a low-transparency 
Quantum Point Contact (QPC) or (which is almost the same) 
by just a tunnel junction [Fig.\ 2(a)]. 
Basically following the model of Ref.\ \cite{Gurvitz-97} we assume 
that the detector barrier height depends on the location of the 
electron in either dot 1 or 2; then the current through 
the tunnel junction (which is the detector output) is sensitive 
to the electron location. 

The Hamiltonian of the system, 
        \begin{equation} 
{\cal H} = {\cal H}_{QB}+{\cal H}_{DET} +{\cal H}_{INT},
        \end{equation}
consists of terms describing the double-dot qubit, the detector, 
and their interaction. 
The qubit Hamiltonian, 
        \begin{equation}
{\cal H}_{QB} = \frac{\varepsilon}{2}\, (c_2^\dagger  c_2 - c_1^\dagger c_1)
        + H \, (c_1^\dagger c_2 +c_2^\dagger c_1) , 
        \label{QB}\end{equation} 
is characterized by the energy asymmetry $\varepsilon$  between two 
dots and the tunneling strength $H$ 
(we assume real $H$ without loss of generality).
The detector and interaction 
Hamiltonians can be written as 
        \begin{eqnarray}
&& {\cal H}_{DET} = \sum_l E_l a_l^\dagger a_l +\sum_r E_r a_r^\dagger a_r 
+ \sum_{l,r} T (a_r^\dagger a_l+ a_l^\dagger a_r),  
        \nonumber\\ 
&& {\cal H}_{INT}= \sum_{l,r} \frac{\Delta T}{2} \, 
(c_1^\dagger c_1 - c_2^\dagger c_2) 
(a_r^\dagger a_l + a_l^\dagger a_r),
        \end{eqnarray}
where both $T$ and $\Delta T$ are assumed real and their dependence on 
the states in electrodes ($l,r$) is neglected. For simplicity we assume
zero temperature (Bayesian formalism at finite temperatures has been
considered in Refs.\ \cite{Kor-rev,Goan,Kor-osc,Rus-osc}). 
If the electron occupies dot 1, then the average current through
the detector is $I_1= 2\pi (T+\Delta T/2)^2 \rho_l\rho_r e^2V/\hbar$ 
($V$ is the voltage across the tunnel junction and $\rho_{l,r}$ are 
the densities of states in the electrodes), while if the measured electron 
is in the dot 2, the average current is 
$I_2=2\pi (T-\Delta T/2)^2 \rho_l\rho_r e^2V/\hbar$.

        The difference between the currents, 
        \begin{equation}
\Delta I\equiv I_1-I_2,  
        \end{equation} 
determines the detector response to the electron position.
    Because of the detector shot noise, the two 
states cannot be distinguished instantaneously and 
the signal-to-noise ratio (S/N) gradually improves with the increase of 
the measurement duration. The S/N becomes close to unity after
the ``measurement''  time  
        \begin{equation}
\tau_m = \frac{(\sqrt{S_1} +\sqrt{S_2})^2}{2(\Delta I)^2}, 
        \label{tau_m}\end{equation} 
where the spectral densities $S_1$ and $S_2$ of the 
detector shot noise for states $|1\rangle$ and $|2\rangle$ are given 
by the Schottky formula,  
        \begin{equation}
S_{1,2}=2eI_{1,2}.
        \label{Schottky} \end{equation} 
[Actually, Eq.\ (\ref{tau_m}) exactly corresponds to S/N=1 for $S_1=S_2$,
while for S/N$\neq$1 it gives the proper asymptotic scaling at 
$t\gg \tau_m$.]
To avoid an explicit account of the detector quantum noise we will 
consider only processes at frequencies $\omega \ll eV/\hbar$; 
in particular, we assume $\max (\hbar\tau_m^{-1}, |\varepsilon|, |H|)  
\ll eV$. 

Notice that due to electron charge discreteness and stochastic nature 
of tunneling, the total number $n(t)$ of electrons passed through the 
detector is sometimes a more convenient magnitude to work with than 
the current $I(t)=e\,dn(t)/dt$. In particular, we will use $n(t)$ instead
of $I(t)$ for the Bayesian formalism derivation in the next section.

\section{Derivation of the Bayesian formalism via ``Bloch'' equations} 

        The ``conventional'' ensemble-averaged equations for the qubit
density matrix $\rho_{ij} (t)$, 
        \begin{eqnarray}
&&\dot{\rho}_{11}=  -\dot{\rho}_{22}=  -2\, \frac{H}{\hbar} \,
        \mbox{Im}\, \rho_{12} , 
        \label{conv1}\\
&& {\dot\rho}_{12}=  \imat \, \frac{\varepsilon}{\hbar} \,\rho_{12}+ 
        \imat \,\frac{H}{\hbar } \,
(\rho_{11}-\rho_{22}) -\Gamma_d \, \rho_{12},  
        \label{conv2}\end{eqnarray}
do not take into account any information about the detector outcome 
and describe the effect of continuous measurement by the ensemble 
decoherence rate \cite{Gurvitz-97}
        \begin{equation}
\Gamma_d = \frac{(\sqrt{I_1}-\sqrt{I_2})^2}{2e}. 
        \label{GdGur}\end{equation}
(Notice a relation  
$\Gamma_d \tau_m =1/2$; 
as will be seen later, this means that the detector is ideal). 

        Equations (\ref{conv1})--(\ref{conv2}) imply tracing over all
detector degrees of freedom, including the measurement outcome. 
An important step towards taking into 
account the measurement record was a derivation \cite{Gurvitz-97} 
of ``Bloch'' equations for the density matrix $\rho_{ij}^n(t)$ which 
is divided into components with different number 
$n$ of electrons passed through the detector: 
        \begin{eqnarray} 
&& \dot\rho_{11}^{\,n} = 
        - \frac{I_1}{e} \, \rho_{11}^n + \frac{I_1}{e}\, \rho_{11}^{n-1} 
        -2\, \frac{H}{\hbar} \, \mbox{Im}\, \rho_{12}^n \, ,
        \label{Gur1}\\ 
&& \dot\rho_{22}^{\,n} = 
        - \frac{I_2}{e} \, \rho_{22}^n + \frac{I_2}{e}\, \rho_{22}^{n-1}  
        + 2\, \frac{H}{\hbar} \, \mbox{Im}\, \rho_{12}^n  \, , 
        \label{Gur2}\\ 
&& {\dot\rho}_{12}^{\, n} = 
        \imat \, \frac{\varepsilon}{\hbar} \,\rho_{12}^n+ 
        \imat \,\frac{H}{\hbar } \,(\rho_{11}^n-\rho_{22}^n) 
        -\frac{I_1+I_2}{2e} \, \rho_{12}^n    
\nonumber\\
&&\hspace{1cm}   + \frac{\sqrt{I_1I_2}}{e} \, \rho_{12}^{n-1} \, .   
        \label{Gur3}\end{eqnarray}
Eqs.\ (\ref{conv1})--(\ref{GdGur}) can be obtained from the Bloch equations
using summation over $n$ and relation $\rho_{ij}=\sum_n \rho_{ij}^n$.
(Absence of nondiagonal in $n$ matrix elements $\rho_{ij}^{nm}$ is related
to the assumption of large detector voltage \cite{Gurvitz-97}.) 

     Despite the Bloch equations carry the total number $n$ of electrons
passed through the detector, they cannot take into account the whole 
measurement record $n(t)$ for a particular realization of measurement
process. We should make a simple but important step for that: 
{\it we should introduce a sufficiently frequent collapse of $n(t)$ 
corresponding to a particular realization of the measurement 
record} \cite{Kor-rev}. This idea is illustrated 
in Fig.\ 2(b). Including ``detector'' into the quantum part of the setup,
we anyway have to deal with a classical information, so we introduce
a classical ``pointer'' which periodically (at times $t_k$) forces the 
system ``qubit+detector'' to choose a definite value for $n(t_k)$. 
An obvious drawback of such construction is that it is absolutely not clear 
what should be a sequence of $t_k$ (in other words, how strongly the 
detector
and pointer should be coupled). In general, the frequency of this collapse
can depend on the physical parameters of interaction between the 
measurement stage included in the ``detector'' Hamiltonian and the next 
stage. 
The only obvious fact is that if in an experiment we can {\it record} 
$n(t)$ with some frequency, then the collapse should take place at least 
not less frequent. Still it is unclear how much more frequent. 
Fortunately, for a model described in the previous section the results 
do not depend on the choice of $t_k$ if $\Delta t_k \equiv t_k-t_{k-1}$
are sufficiently small, so the natural choice is 
$\Delta t_k \rightarrow 0$.

        Technically the procedure is the following. During the time 
between $t_{k-1}$ and $t_k$ the ``qubit+detector'' evolves according 
to the Bloch equations (\ref{Gur1})--(\ref{Gur3}), while at time $t_k$ 
the number $n$ is collapsed in the ``orthodox'' way \cite{Neumann}.
This means that the probability $P(n)$ of getting some $n(t_k)$
is equal to 
        \begin{equation}
P(n) = \rho_{11}^{n}(t_k) +\rho_{22}^{n}(t_k), 
        \label{P(n)}\end{equation}
while after a particular $n_k$ is picked, the density matrix $\rho_{ij}^n$
is immediately updated (collapsed): 
        \begin{eqnarray}
&& \rho_{ij}^n(t_k+0) =\delta_{n,n_k} \, \rho_{ij}(t_k+0), 
        \label{collapse1}\\ 
&& \rho_{ij}(t_k+0)= \frac{\rho_{ij}^{n_k}(t_k-0)} 
{ \rho_{11}^{n_k}(t_k-0) +\rho_{22}^{n_k}(t_k-0) } , 
        \label{collapse2}\end{eqnarray} 
where $\delta_{n,n_k}$ is the Kronecker symbol.  
After that the evolution is again described by 
Eqs.\ (\ref{Gur1})--(\ref{Gur3}) with $n$ shifted by $n_k$ 
until the next collapse occurs at $t=t_{k+1}$, and so on.
{\it This procedure is the main step} in the derivation of the Bayesian
formalism. 

        Let us discuss now the relation of this procedure to the
classical Bayes theorem \cite{Bayes,Borel} which says that
a posteriori probability $P(B_i|A)$ of a hypothesis $B_i$ after 
learning an information $A$ ($B_i$ form a complete set of 
mutually exclusive hypotheses) is equal to 
        \begin{equation}
P(B_i|A)=\frac{P(B_i)P(A|B_i)}{\sum_k P(B_k)P(A|B_k)}
        \label{Bayes-cl}\end{equation} 
where $P(B_i)$ is a priori probability of the hypothesis $B_i$ 
(before learning information $A$) and $P(A|B_i)$ is the 
conditional probability of the event $A$ under hypothesis $B_i$. 

        Assuming for a moment $H=0$ and $\varepsilon =0$ in the qubit 
Hamiltonian 
(so that the qubit evolution is due to measurement only), it is easy
to find \cite{Kor-rev} that Eqs.\ (\ref{Gur1})--(\ref{Gur3}) 
and our procedure (\ref{collapse1})--(\ref{collapse2}) lead 
to the qubit evolution as 
        \begin{eqnarray}
&& \rho_{11}(t_k) = \frac{\rho_{11}(t_{k-1}) P_1(\Delta n_k)}
 {\rho_{11}(t_{k-1}) P_1(\Delta n_k)+\rho_{22}(t_{k-1}) P_2(\Delta n_k)},
\label{r11tk}\\
&& \rho_{22}(t_k) = \frac{\rho_{22}(t_{k-1}) P_2(\Delta n_k)} 
 {\rho_{11}(t_{k-1}) P_1(\Delta n_k)+\rho_{22}(t_{k-1}) P_2(\Delta n_k)},
\label{r22tk}\\
&& \rho_{12}(t_k)=\rho_{12}(t_{k-1}) 
        \frac{[\rho_{11}(t_k)\rho_{22}(t_k)]^{1/2}}
             {[\rho_{11}(t_{k-1})\rho_{22}(t_{k-1})]^{1/2}} ,
        \label{r12tk}\end{eqnarray}
where $\Delta n_k = n_k-n_{k-1}$ is the number of electrons passed 
through the detector during time $\Delta t_k$ and
        \begin{equation}
P_i(n)=\frac{(I_i\Delta t_k/e)^{n}}{n!} \, 
\exp (-I_i\Delta t_k/e)
 	\label{Poisson}   \end{equation}
is the classical Poisson distribution for this number assuming either qubit 
state $|1\rangle$ or $|2\rangle$. One can see that the diagonal matrix 
elements $\rho_{ii}$  
{\it exactly obey the classical Bayes formula} (\ref{Bayes-cl}),
i.e.\ {\it as if} the qubit is really either in the state $|1\rangle$ or 
$|2\rangle$, but not in both simultaneously. 
Actually, this fact is not much surprising because at least in some 
sense 
$\rho_{ii}$ are the probabilities. 
Equation (\ref{r12tk}) is a little more surprising and says that the
measurement preserves the degree of qubit purity $\rho_{12}/(\rho_{11}
\rho_{22})^{1/2}$; for instance, {\it a pure state remains pure during the
whole measurement process}. 

        After introducing the main procedure 
(\ref{collapse1})--(\ref{collapse2}),
further derivation of the Bayesian formalism is pretty simple
and depends on whether we want to consider finite detector response,
$|\Delta I| \sim I_0\equiv (I_1+I_2)/2$ or a weak response, 
$|\Delta I| \ll I_0$. 
In the first case each event of tunneling through the detector carries
significant information and significantly affects the qubit state, 
so a reasonable ``experimental'' setup implies
recording the time of each tunneling event. 
Then during the time periods when no electrons are passing through the 
detector, the evolution is essentially described by the Bloch equations 
(\ref{Gur1})--(\ref{Gur3}) with $n=0$, while the frequent collapses
[$\Delta t_k \ll \min (e/I_1, e/I_2, \hbar /H, \hbar /\varepsilon )$] just
restore the density matrix normalization, leading to the continuous
(but not unitary!)  qubit evolution \cite{Kor-rev,Goan}: 
        \begin{eqnarray}
&& \dot\rho_{11} = -\dot\rho_{22}= 
-2\,\frac{H}{\hbar}\,\mbox{Im}\, \rho_{12} 
-\frac{\Delta I}{e}\, \rho_{11}\rho_{22} , 
        \label{QJ1}\\
&& \dot\rho_{12} =
\frac{\imat\varepsilon}{\hbar}\, \rho_{12} 
+\frac{\imat H}{\hbar} \, (\rho_{11}-\rho_{22})
+ \frac{\Delta I}{2e} (\rho_{11}-\rho_{22}) \rho_{12} .
        \label{QJ2}\end{eqnarray}
However, at moments when one electron passes through the detector, the
qubit state changes abruptly (corresponding to $\Delta n_k=1$ and
$\Delta t_k\rightarrow 0$ in Eqs.\ (\ref{r11tk})--(\ref{r12tk})): 
        \begin{eqnarray}
&& \rho_{11}(t+0) =\frac{I_1\rho_{11}(t-0)}
{I_1\rho_{11}(t-0)+I_2\rho_{22}(t-0)} \, , 
        \label{QJ3}\\
&& \rho_{22}(t+0)=1-\rho_{11}(t+0),
        \label{QJ4}\\ 
&& \rho_{12}(t+0)= \rho_{12}(t-0) \left[ \frac{\rho_{11}(t+0)\,
\rho_{22}(t+0)}
{\rho_{11}(t-0)\,\rho_{22}(t-0)} \right] ^{1/2}.  
        \label{QJ5}\end{eqnarray}
These abrupt changes are usually called ``quantum jumps'' \cite{Goan}.
Notice that for $I_1 > I_2$ the jumps always shift the qubit state closer
to $|1\rangle$ (because detector tunneling is ``more likely'' for state
$|1\rangle$), while continuous nonunitary evolution shifts the state
towards $|2\rangle$. On average the evolution is still given by 
conventional Eqs.\ (\ref{conv1})--(\ref{GdGur}). 

        The case of a weak detector response, $|\Delta I| \ll I_0$,
is more realistic from the experimental point of view. In this case 
the measurement time $\tau_m$ as well as the ensemble decoherence time
$\Gamma_d^{-1}$ are much longer than the average time $e/I_0$ between
electron passages in the detector. If the tunneling in the qubit is
also sufficiently slow, $\hbar /H \gg e/I_0$, we can completely disregard 
individual events in the detector and consider the detector current $I(t)$ 
as quasicontinuous. 
Then Eqs.\ (\ref{r11tk})--(\ref{r22tk}) for the evolution
due to measurement only (neglecting $H$ and $\varepsilon$) transform into
equations which again have clear Bayesian interpretation: 
        \begin{eqnarray}
 \rho_{11}(t+\tau ) = 
\frac{\rho_{11}(t) P_1({\overline I})}
{\rho_{11}(t) P_1({\overline I})+\rho_{22}(t) P_2({\overline I})} ,
        \label{r11tt}\\
 \rho_{22}(t+\tau ) = 
\frac{\rho_{22}(t) P_2({\overline I})}
{\rho_{11}(t) P_1({\overline I})+\rho_{22}(t) P_2({\overline I})} ,
        \label{r22tt}\end{eqnarray}
where 
        \begin{equation}
\overline{I}\equiv \frac{1}{\tau}\int_t^{t+\tau} I(t' ) \, dt'
        \label{Iav-def}\end{equation}
is the detector current averaged over the time interval $(t, t+\tau )$
and $P_i({\overline I})$ are its classical Gaussian probability 
distributions for two qubit states: 
        \begin{equation}
P_i(\overline{I}) =   \frac{1}{(2\pi D)^{1/2}} \, 
\exp [-\frac{(\overline{I}-I_i)^2}{2D}] , \,\,\, 
D=S_0/2\tau , 
        \label{P_i}\end{equation} 
(here $S_0=2eI_0$ is the detector noise), 
while Eq.\ (\ref{r12tk}) essentially does not change. 

        Differentiating Eqs. (\ref{r11tt}), (\ref{r22tt}), and 
(\ref{r12tk})
over time and including additional evolution due to $H$ and $\varepsilon$,
we obtain the {\it main equations of the Bayesian formalism}: 
        \begin{eqnarray}
&&  \dot{\rho}_{11}=  -\dot{\rho}_{22}=  
-2\,\frac{H}{\hbar}\,\mbox{Im}\,\rho_{12}
         +\rho_{11}\rho_{22}\, \frac{2\Delta I}{S_0}\, [I(t)-I_0], 
        \label{Bayes1}\\ 
&&  {\dot\rho}_{12}=  \imat\, \frac{\varepsilon}{\hbar }\,
	\rho_{12}+ 
        \imat \, \frac{H}{\hbar } \, (\rho_{11}-\rho_{22})
\nonumber \\
&& \hspace{1cm}  -( \rho_{11}-  \rho_{22})  \frac{\Delta I}{S_0} \, 
[I(t)-I_0]\, \rho_{12} .
        \label{Bayes2} \end{eqnarray}
In each realization of measurement the noisy detector outcome $I(t)$ is
different; however, for each realization we can precisely monitor the
evolution of the qubit density matrix, plugging experimental $I(t)$
into Eqs.\ (\ref{Bayes1})--(\ref{Bayes2}). Let us stress again that these 
equations show the absence of any qubit decoherence during the process 
of measurement. Pure initial state remains pure; moreover, initially 
mixed state gradually purifies during the measurement
if $H\neq 0$ \cite{Kor-rev,Kor-99}.
The gradual state purification is essentially due to acquiring more
and more information about the qubit state from the measurement record. 

        While the qubit state does not decohere in each individual 
realization of the measurement, different members of the ensemble evolve
differently because of random $I(t)$. Averaging Eqs.\ 
(\ref{Bayes1})--(\ref{Bayes2}) over random $I(t)$ and using the relation 
[which follows from Eqs.\ (\ref{P(n)}), (\ref{r11tk})--(\ref{r22tk}), and 
(\ref{Poisson})] 
        \begin{equation}
I(t) -I_0 = \frac{\Delta I}{2}\, (\rho_{11} -\rho_{22}) +\xi (t) ,
        \label{I(t)}\end{equation}
where $\xi (t)$ is a zero-correlated (``white'') random process with 
the same spectral density as the detector noise, $S_\xi =S_0$,
we obtain conventional Eqs.\ (\ref{conv1})--(\ref{conv2}). Therefore,
the ensemble-averaged decoherence in our model is just a consequence
of averaging over the measurement outcome (similar conclusion is also
valid in the finite response case).

        Notice that since $I(t)$ contains the white noise contribution, 
Eqs.\ (\ref{Bayes1})--(\ref{Bayes2}) are  nonlinear stochastic differential 
equations \cite{Oksendal} and dealing with them requires a special care.
The problem is that their analysis depends on the choice of the derivative 
definition. Two mainly used definitions are the symmetric 
derivative: $\dot \rho (t) \equiv \lim_{\tau\rightarrow 0}
[\rho (t+\tau /2)-\rho (t-\tau /2)]/\tau$ which leads to the so-called
Stratonovich interpretation of the stochastic differential equations, and 
the forward derivative: $\dot \rho (t) \equiv \lim_{\tau\rightarrow 0} 
[\rho (t+\tau )- \rho (t)]/\tau $ (It\^o interpretation). 
Usual calculus rules remain valid only in the Stratonovich form 
\cite{Oksendal}, so the 
physical intuition works better when using Stratonovich definition.
However, It\^o interpretation allows simple averaging over the noise
and because of that is usually preferred by mathematicians. 
Since we derived Eqs.\ (\ref{Bayes1})--(\ref{Bayes2}) by a simple 
first-order 
differentiation, we automatically obtained them in the Stratonovich
form (keeping the second-order terms in the expansion, we can
obtain different equations, depending on the definition of the derivative).
Since sometimes It\^o form is more preferable, let us translate Bayesian
equations into It\^o form using the following general rule \cite{Oksendal}. 
For an arbitrary system of equations 
        \begin{equation} 
\dot x_i (t) = G_i({\bf x} , t) + F_i({\bf x} ,t )\, \xi (t) 
        \end{equation} 
in Stratonovich interpretation, the corresponding It\^o equation
which has the same solution is 
        \begin{equation}
\dot x_i(t) = G_i({\bf x} , t) 
        +  \frac{S_\xi }{4} \sum_k 
        \frac{\partial F_i({\bf x} ,t)}{\partial x_k} \, F_k({\bf x} ,t)
        + F_i({\bf x} , t)\, \xi (t)  \, , 
        \end{equation} 
where $x_i(t)$ are the components of the vector ${\bf x}(t)$, \,\, 
$G_i$ and $F_i$ 
are arbitrary functions, and 
the constant $S_\xi $ is the spectral density of the white noise process 
$\xi (t)$. Applying this transformation to Eqs.\ 
(\ref{Bayes1})--(\ref{Bayes2}), 
 we get the following equations in It\^o interpretation:
        \begin{eqnarray}
&& 
   \dot{\rho}_{11}  =   -\dot{\rho}_{22}=  -2\,\frac{H}{\hbar}\,\mbox{Im}\,
          \rho_{12}
         +\rho_{11}\rho_{22}\, \frac{2\Delta I}{S_0}\, \xi (t)\, ,  
        \label{Ito1}\\ 
&&
 {\dot\rho}_{12}  =   \imat\, \frac{\varepsilon}{\hbar }\,\rho_{12} +   
        \imat \, \frac{H}{\hbar } \, (\rho_{11}-\rho_{22})
\nonumber\\
&& \hspace{1cm}
 -  ( \rho_{11}-  \rho_{22})  \frac{\Delta I}{S_0} \, 
\rho_{12}\, \xi (t) - \frac{(\Delta I)^2}{4S_0}  \, 
\rho_{12} \, , 
        \label{Ito2} 
        \end{eqnarray}
while the relation between pure noise $\xi (t)$ and the current $I(t)$ 
is still given by Eq.\ (\ref{I(t)}). Notice that the last term in
Eq.\ (\ref{Ito2}) does not actually mean the single qubit decoherence 
(pure state remains pure), 
but is just a feature of the It\^o form [it directly corresponds to the 
ensemble decoherence after averaging over $\xi(t)$].

\section{Derivation of the formalism via correspondence principle} 

        Another derivation \cite{Kor-99} of the Bayesian formalism 
for a single qubit can be based on the logical use of the correspondence
principle \cite{Neumann}, classical Bayes formula, and results of the 
conventional ensemble-averaged formalism. 
Even though this way lacks some advantages of the ``microscopic'' 
derivation 
discussed in the previous section, it can be applied to a broader class
of solid-state detectors, in particular, to the finite-transparency
quantum point contact and (with some extension) to the single-electron
transistor and SQUID. In this section we will assume a double-dot qubit
measured by a finite-transparency QPC and treat the detector current $I(t)$
as a  quasicontinuous  noisy  signal  that  implies  weak  detector 
response,
 $|\Delta I|\ll I_0$. 

        Let us start with a completely {\it classical} case when 
the electron is actually located in one of two dots and does not move, 
but we do not 
know in which one, so the measurement gradually reveals the actual electron
location. This is a well studied problem of the probability theory. 
The measurement process can be described as an evolution 
of probabilities (we still call them $\rho_{11}$ and $\rho_{22}$) 
which reflect our knowledge about the system state. 
 Then for arbitrary $\tau$ (which can be comparable to $\tau_m$) 
the current $\overline{I}$ averaged over time interval $(t,t+\tau)$ 
[see Eq.\ (\ref{Iav-def})] has the probability distribution 
          \begin{equation}
P({\overline I}) = \rho_{11}(t) P_1({\overline I}) +
\rho_{22}(t) P_2({\overline I}) , 
        \label{P(I)}\end{equation}
where  $P_i$ are given by Eq.\ (\ref{P_i}) and depend on the detector 
white noise spectral density $S_0$ which should not necessarily satisfy
Schottky formula. After the measurement during time 
$\tau$ the information about the system state has increased 
and the probabilities $\rho_{11}$ and $\rho_{22}$ should be updated using 
the measurement result $\overline{I}$ and the Bayes formula 
(\ref{r11tt})--(\ref{r22tt}). This completely describes the classical 
measurement process. 

        The next step in the derivation is an important assumption: 
in the quantum case with $H=0$ the evolution of 
$\rho_{11}$ and $\rho_{22}$ is still given by Eqs.\ 
(\ref{r11tt})--(\ref{r22tt}) 
because there is no possibility to distinguish between classical
and quantum cases, performing only this kind of measurement. Even though
this assumption is quite obvious, it is not derived formally but should 
rather be regarded as a {\it consequence of the correspondence principle}.
 
The correspondence with classical measurement cannot describe the evolution
of $\rho_{12}$; however, there is still an upper limit: $|\rho_{12}| 
\leq [\rho_{11}\rho_{22}]^{1/2}$. Surprisingly, this inequality 
is sufficient for exact calculation of $\rho_{12}(t)$ in the 
case of a QPC as a detector (we still assume $H=0$). 
Averaging the inequality over all possible detector 
outputs $\overline{I}$ using distribution (\ref{P(I)}) we get the 
inequality 
        \begin{eqnarray}
&&  |\langle \rho_{12}(t+\tau )\rangle | \leq 
 \langle |\rho_{12}(t+\tau )| \rangle  \leq 
\langle [\rho_{11} (t+\tau )\rho_{22} (t+\tau )]^{1/2} \rangle 
	\nonumber \\ 
&& \hspace{2cm}  = [\rho_{11}(t)\rho_{22}(t)]^{1/2} 
\exp [-(\Delta I)^2\tau /4S_0] 
        \label{inequality}\end{eqnarray} 
[here the decaying exponent is a consequence of changing $\rho_{11}$ and 
$\rho_{22}$ that reduces their average product]. 
On the other hand, from the conventional approach we know 
\cite{Averin,Hackenbroich,Aleiner,Kor-Av} 
that the ensemble-averaged qubit decoherence rate caused by a QPC is equal
to $\Gamma_d=(\Delta I)^2/4S_0$, where $S_0=2eI_0(1-{\cal T})$ and 
${\cal T}$ 
is the QPC transparency. This means that inequality (\ref{inequality}) 
actually {\it reaches its upper bound}.
This is possible {\it only if\/} in each realization of the measurement 
process an initially pure density matrix $\rho_{ij} (t)$ 
stays pure all the time, $|\rho_{12}|^2=\rho_{11}\rho_{22}$. 
Moreover, since the phase of $\rho_{12}(t+\tau )$ should be the same
for all realizations (to ensure $|\langle \rho_{12}(t+\tau )\rangle | 
= \langle |\rho_{12}(t+\tau )|\rangle$), 
the only possibility in absence of a detector-induced 
shift of $\varepsilon$ is
        \begin{equation}
\frac{\rho_{12}(t+\tau )}{[\rho_{11}(t+\tau )\rho_{22}(t+\tau )]^{1/2}}=
\frac{\rho_{12}(t)}{[\rho_{11}(t)\rho_{22}(t)]^{1/2}} \, 
\mbox{e}^{\imat \varepsilon \tau /\hbar } 
        \label{r12tt}\end{equation}
(if the coupling with detector shifts $\varepsilon$, we just have to use
the shifted value).

        As the next step of the derivation, let us allow an arbitrary mixed 
initial state of the qubit (but still $H=0$). It can always be represented 
as a mixture of two states with the same diagonal matrix elements, one of
which is pure, while the other state does not have nondiagonal matrix 
elements. Since nondiagonal matrix elements for the latter state cannot 
appear in the process of measurement and since the evolution of the 
diagonal matrix elements is equal for both states, one can conclude that
Eq.\ (\ref{r12tt}) remains valid, i.e.\ for mixed states the degree 
of purity is conserved (gradual purification does not occur at $H=0$). 
        The final step of the formalism derivation is differentiating 
Eqs.\ (\ref{r11tt}), (\ref{r22tt}), and (\ref{r12tt}) over time
and adding (in a simple way) the evolution due to $H$. 

In this way we  reproduce Eqs.\ (\ref{Bayes1})--(\ref{Bayes2}).
However, as seen from the derivation, now they are applicable to a 
{\it broader class of detectors} (which includes the finite-transparency 
QPC) for which $\Gamma_d=(\Delta I)^2/4S_0$. This relation can also be
expressed as $\Gamma_d\tau_m=1/2$ since $\tau_m =2S_0/(\Delta I)^2$ 
for a weakly responding detector. 
As will be discussed in the next section, this is a condition of 
an ideal quantum detector.

\section{Nonideal detectors} 
        
        The relation $\Gamma_d\tau_m =1/2$ which is valid for
the models of a tunnel junction and QPC as detectors, basically says 
that the {\it larger output noise} $S_0$ {\it of a detector leads to a 
smaller backaction} characterized by ensemble decoherence $\Gamma_d$. 
This is quite expected from quantum mechanical point of view (the faster 
we get information, the faster we should collapse the measured state).
However, it is obviously not necessarily the case for an arbitrary 
solid-state detector; for example, the increase of output noise 
$S_0$ can be due to later stages of signal amplification, which do not
affect $\Gamma_d$. In other words, it is easy to imagine a bad detector
which produces a lot of both output and backaction noises. 

        To take into account an extra detector noise, we can 
phenomenologically add a dephasing rate $\gamma_d$ into the Bayesian 
equations: 
        \begin{eqnarray}
&&  \dot{\rho}_{11}=  -\dot{\rho}_{22}=  
-2\,\frac{H}{\hbar}\,\mbox{Im}\,\rho_{12}
         +\rho_{11}\rho_{22}\, \frac{2\Delta I}{S_0}\, [I(t)-I_0],  
        \label{Bayes1m}\\  
&& 
{\dot\rho}_{12}=  \imat\, \frac{\varepsilon}{\hbar }\,\rho_{12}+ 
        \imat \, \frac{H}{\hbar } \, (\rho_{11}-\rho_{22})
\nonumber \\
&&\hspace{1cm}
  -( \rho_{11}-  \rho_{22})  \frac{\Delta I}{S_0} \, 
[I(t)-I_0]\, \rho_{12} -\gamma_d \rho_{12}.
        \label{Bayes2m}\end{eqnarray}
This obviously increases the ensemble decoherence rate:
        \begin{equation}
        \Gamma_d = \frac{(\Delta I)^2}{4S_0}+\gamma_d . 
        \label{Gamma-gamma}\end{equation} 
A natural definition of a detector ideality (quantum efficiency) in this 
case is
        \begin{equation}
\eta \equiv 1- \frac{\gamma_d}{\Gamma_d}=\frac{1}{2\Gamma_d\tau_m}. 
        \end{equation}
An upper limit for $\eta$ is 100\% because of a fundamental 
limitation 
        \begin{equation}
\Gamma_d\tau_m \geq 1/2,
        \label{limitation}\end{equation}
which is a by-product of the Bayesian derivation for the case of 
quasicontinuous detector current and small difference between noises 
$S_1$ and $S_2$ (so that $S_1=S_2=S_0$) -- see inequality (\ref{inequality}). 
[In the case of a detector with $S_1\neq S_2$ and possibility to observe 
each passing electron, Eq.\ (\ref{limitation}) remains valid; however, 
a meaningful model would imply Poisson statistics (\ref{Poisson}) 
and therefore Schottky 
formula for the detector noise.]  

        The extra dephasing $\gamma_d$ in Eq.\ (\ref{Bayes2m}) can be 
interpreted \cite{Kor-99} as an effect of extra environment or 
(mathematically) as due to a second detector ``in parallel'', the 
output of which is not read out (then we have to average over possible
outputs). It can be also interpreted as an effect of extra noise $S_{add}$ 
at the detector output, $S_0=(\Delta I)^2/4\Gamma_d+S_{add}$. 
In this latter case one can argue that the qubit 
is actually in a pure state and the evolution of the diagonal matrix 
elements is actually different from what is given by Eq.\ (\ref{Bayes1m}),
because the measured current $I(t)$ is not the ``actual'' detector current.
Yes, we would know the exact pure state if our amplifiers did not produce
extra noise $S_{add}$; however, since we do not have access to the 
``actual'' 
detector current, we should perform averaging over the extra noise.
It is easy to show that such averaged qubit density matrix (which
is a density matrix ``for us'') still satisfy Eqs.\ 
(\ref{Bayes1m})--(\ref{Bayes2m}). 

        Introduction of the detector ideality $\eta$ allows us to consider 
a continuous transition from the conventional ensemble-averaged result 
(\ref{conv1})--(\ref{conv2}) to the ``pure'' Bayesian result 
(\ref{Bayes1})--(\ref{Bayes2}). The effect of a pure environment can be 
considered as a measurement with an extremely bad detector, $\eta =0$. 
Technically it corresponds to $\Delta I=0$ in Eqs.\ 
(\ref{Bayes1m})--(\ref{Bayes2m}), transforming them into conventional 
equations. 
The case of a detector with very small efficiency, $\eta \ll 1$, can 
be treated in two steps: first, we analyze the effect of the decoherence
term ($\gamma_d \approx \Gamma_d$), and then we use the classical (still
Bayesian) analysis to relate the qubit density matrix and the measurement
outcome. So, only for good detectors with the efficiency $\eta$ comparable
to unity, the quantum Bayesian approach discussed in this paper is really
necessary. Some people could argue that it is so difficult to create 
a solid-state 
detector with good quantum efficiency, that the Bayesian formalism
is useless at the present-day level of technology. However, actually at 
present such detectors are becoming available. For example, the analysis 
of experimental data of the recent ``which path'' experiment \cite{Buks} 
shows that their QPC had a quantum efficiency quite close to 100\%.

        The account of the detector nonideality by introducing extra
decoherence rate $\gamma_d$ into Eq.\ (\ref{Bayes2m}) implicitly assumes
the absence of a direct correlation between the output detector noise 
and the backaction noise affecting the qubit energy asymmetry 
$\varepsilon$. However, such correlation is a typical situation, for 
example, for a single-electron transistor as a detector \cite{Kor-noise}. 
In this case the knowledge of the noisy detector output $I(t)$ 
gives some information about the probable backaction noise ``trajectory'' 
$\varepsilon (t)$ which can be used to improve our knowledge of the 
qubit state. Compensation for the most probable trajectory 
$\varepsilon (t)$ leads to the improved Bayesian equations for the SET
in which Eq.\ (\ref{Bayes2m}) is replaced with 
        \begin{eqnarray} 
&& {\dot\rho}_{12}=  \imat\, \frac{\varepsilon}{\hbar }\,\rho_{12}+ 
        \imat \, \frac{H}{\hbar } \, (\rho_{11}-\rho_{22})
\nonumber \\
&& \hspace{1cm}
-( \rho_{11}-  \rho_{22})  \frac{\Delta I}{S_0} \, 
[I(t)-I_0]\, \rho_{12} 
        \nonumber \\
&& \hspace{1cm} +\, \imat \, K\, [I(t)-I_0] \, \rho_{12} 
 -\tilde\gamma_d \rho_{12},     
        \label{Bayes2mm} 
        \end{eqnarray}
where $K=(d\varepsilon /d\varphi )S_{I\varphi }/S_0\hbar$, $\varphi$ is
the electric potential of the SET central electrode, and $S_{I\varphi }$ 
is the mutual low-frequency spectral density between the 
current noise  and $\varphi$ noise. 
[Strictly speaking, $S_{I\varphi}$ in our notation is only the usual real 
part of the mutual spectral density, which reflects the detector 
``asymmetry'', while the imaginary part can formally describe the detector 
response \cite{Averin}. 
Also notice that since the small shift of the SET operating point for two 
localized 
qubit states in general affects the energy $\varepsilon$, it should be 
defined self-consistently in Eq.\ (\ref{Bayes2mm}).] 

The dephasing rate $\tilde \gamma_d $ should now satisfy equation 
        \begin{equation} 
\tilde\gamma_d =\Gamma_d - \frac{(\Delta I)^2}{4S_0} - \frac{K^2S_0}{4} 
\, 
        \end{equation}
to correspond to the the ensemble-averaged 
dynamics still described by Eqs.\ (\ref{conv1})--(\ref{conv2}).

        The obvious inequality $\tilde\gamma_d \geq 0$ (in the opposite 
case the relation $|\rho_{12}|^2\leq \rho_{11}\rho_{22}$ would be violated 
in a single realization of measurement) 
imposes a lower bound for the ensemble decoherence rate $\Gamma_d$: 
        \begin{equation}
\Gamma_d \geq \frac{(\Delta I)^2}{4S_0} + \frac{K^2S_0}{4}\, , 
        \label{Gammad}\end{equation} 
which is stronger than the inequality $\Gamma_d\tau_m \geq 1/2$. 

        Inequality (\ref{Gammad}) can be easily expressed in terms of
the energy sensitivity of an SET. Let us define the 
output energy sensitivity as $\epsilon_I \equiv (dI/dq)^{-2} S_I/2C$ 
where $C$ is the total SET island  capacitance, $dI/dq$  is  the 
response to the externally induced charge $q$, and 
we have changed the notation $S_I\equiv S_0$ for a more symmetric look 
of the formulas. Notice that $\epsilon_I$ has the same dimension as 
$\hbar$. Similarly, let us characterize 
the backaction noise intensity by $\epsilon_\varphi \equiv CS_\varphi/2$ 
and the correlation between two noises by the magnitude 
$\epsilon_{I\varphi}\equiv (dI/dq)^{-1}S_{I\varphi}/2$. Since in absence 
of other decoherence sources $\Gamma_d =S_\varphi (C\Delta E/2e\hbar)^2$, 
where $\Delta E$ is the energy coupling between qubit and single-electron
transistor \cite{Kor-rev,Makhlin}, and using also the reciprocity property 
$\Delta q =C\Delta E/e=d\varepsilon /d\varphi$, we can rewrite Eq.\ 
(\ref{Gammad}) as 
        \begin{equation}
\epsilon \equiv 
(\epsilon_I \epsilon_\varphi -\epsilon_{I\varphi}^2)^{1/2} \geq \hbar /2.
        \end{equation}
This is a result known for 20 years \cite{Danilov} for SQUIDs 
(see also \cite{Averin,Devoret,Averin-cotun,Brink,Zorin,Clerk}). 

 When the limit $\epsilon =\hbar /2$ is achieved, the decoherence rate  
        \begin{equation}
\tilde\gamma_d= \frac{(\Delta I)^2}{4S_I}  
\left[\frac{\epsilon_I \epsilon_\varphi -\epsilon_{I\varphi}^2}
{(\hbar /2)^2} -1 \right]
        \end{equation}
in Eq.\ (\ref{Bayes2mm}) vanishes, $\tilde\gamma_d =0$. In this sense 
the detector is ideal, $\tilde\eta =1$, where 
        \begin{equation}
\tilde\eta \equiv 1- \frac{\tilde\gamma_d}{\Gamma_d} = 
\frac{\hbar^2 (dI/dq)^2}{S_IS_\varphi} +  
\frac{(S_{I\varphi})^2}{S_IS_\varphi} \, , 
        \end{equation} 
even though it can be a nonideal detector ($\eta <1$) by the previous 
definition, $\eta =\hbar^2 (dI/dq)^2/S_IS_\varphi $.

 Another possible definition of the detector efficiency in this case is 
        \begin{equation}
\tilde\eta_2 \equiv \frac{(\hbar /2)^2}{\epsilon_I\epsilon_\varphi 
-\epsilon_{I\varphi}^2}= \frac{\hbar^2 (dI/dq)^2}
{S_I S_\varphi-S_{I\varphi}^2}. 
        \end{equation}
Notice a simple relation, 
        \begin{equation} 
\eta =\tilde{\eta}= \tilde\eta_2=
\frac{(\hbar /2)^2}{\epsilon_I\epsilon_\varphi} =
        \frac{1}{2\Gamma_d\tau_m} \, ,  
        \label{eta-eta}\end{equation}
in absence of correlation between 
        noises of $\varphi (t)$ and $I(t)$, 
        $(S_{I\varphi})^2 \ll S_IS_\varphi$.

        Even though Eqs.\ (\ref{Bayes2mm})--(\ref{eta-eta}) were derived
for the SET as a detector, it is rather obvious that they are applicable
to virtually any solid-state detector with continuous output
(for a dc SQUID the current output should obviously be replaced by the
voltage output). 
In particular, the conclusion that reaching the quantum-limited total 
energy sensitivity $\epsilon = \hbar /2$ is equivalent to the detector 
ideality, is quite general. 

        As we already discussed, the tunnel junction and QPC at zero 
temperature (actually, for small temperatures $\beta^{-1} \ll eV$) 
are theoretically ideal quantum detectors. The fact that 
a SQUID can reach the limit of an ideal detector follows from the results
of Ref.\ \cite{Danilov}. A normal state SET is not a good quantum detector
($\eta \ll 1$) at usual operating points above the Coulomb Blockade
threshold \cite{Kor-rev,Makhlin}. However, its quantum efficiency
improves when we go closer to the threshold \cite{Kor-rev,Devoret}
and becomes much better when the operating point is in the cotunneling
range (below the threshold), in which case the limit of an ideal
detector can be achieved \cite{Averin-cotun,Brink}. Superconducting 
SET is generally better than normal SET as a quantum-limited detector 
and can approach 100\% ideality in the supercurrent regime \cite{Zorin}
as well as in the double Josephson-plus-quasiparticle regime \cite{Clerk}
(possibly a threshold of a quasiparticle current is also a good 
operating point in this sense; however, this regime has been studied
only for the current so far \cite{Av-Pekola}, but not for the noise). 
Finally, the resonant-tunneling SET \cite{Averin-res} can reach  
ideality factor $\tilde\eta_2=3/4$ at large bias and complete 
ideality, $\tilde\eta=\tilde\eta_2=1$, in the small-bias limit.

\section{Some experimental predictions}

\subsection{Direct experiments} 

        The Bayesian equations tell us that we can monitor the qubit 
evolution in a single realization of the measurement process using 
the record of the noisy detector output. In particular, for an ideal
detector {\it we can monitor the qubit wavefunction} (except the overall
phase) 
if the initial qubit state is pure or, for a mixed initial state, 
after monitoring for a sufficiently long time so that the gradual 
purification has enough time to produce a practically pure state. 

        This prediction (and hence, the validity of the Bayesian equations)
can in principle be tested experimentally. 
        For example \cite{Kor-99}, let us first prepare the double-dot in 
the symmetric coherent state, $\rho_{11}=\rho_{22}=|\rho_{12}|=1/2$, 
make $H=0$ (raise the barrier), and begin measurement with a QPC 
[Fig.\ 1(b)].  
According to our formalism, after some time $\tau$ 
(the most interesting case is $\tau \sim \tau_{m}$) 
the qubit state remains pure but becomes asymmetric 
($\rho_{11}\neq \rho_{22}$) and 
can be calculated with Eqs.\ (\ref{Bayes1}) and (\ref{Bayes2}). 
To prove this, an experimentalist can use the knowledge of the 
wavefunction to move the electron ``coherently'' into the first dot with 
100\% probability. (Notice that if the qubit is in a mixed state,
no unitary transformation can end up in the state $|1\rangle$ with 
certainty.)
For instance, experimentalist switches off the detector at $t=\tau $, 
reduces the barrier (to get finite $H$), 
and creates the energy difference 
$\varepsilon = [(1-4|\rho_{12}(\tau )|^2)^{1/2}-1] 
H \mbox{Re}\rho_{12} (\tau )/ |\rho_{12}(\tau )|^{2}$; 
then after the time period
$\Delta t =[\pi-\arcsin (\mbox{Im} \rho_{12}(\tau ) \, 
\hbar\Omega/H)]/\Omega$
the ``whole'' electron will be moved into the first dot, 
that can be checked by the detector switched on again. 
[Here $\Omega \equiv (4H^2+\varepsilon^2)^{1/2}/\hbar$ is the frequency of
unperturbed coherent (``Rabi'') oscillations of the qubit.]

        Another experimental idea \cite{Kor-99} is to demonstrate 
the gradual purification of the double-dot density matrix. 
Let us start with a completely mixed (unknown) state 
($\rho_{11}=\rho_{22}=1/2$, $\rho_{12}=0$) 
of the double-dot qubit with finite $H$.  
Then using the detector output $I(t)$ and Eqs.\ 
(\ref{Bayes1})--(\ref{Bayes2}) an experimentalist 
gradually gets more and more knowledge about the randomly evolving 
qubit state (gradual purification), 
eventually ending up with almost pure wavefunction with 
precisely known {\it phase} of Rabi oscillations
(we are not talking about the wavefunction phase,
but about the phase of diagonal matrix elements oscillations). 
The final check of the wavefunction can be similar to that 
described in the previous paragraph. It can be even simpler, 
since with the knowledge of the phase of oscillations 
it is easy to stop the evolution by raising the barrier when the 
electron is in the first dot with certainty. Notice that if fast real-time 
calculations are not available, the moment of raising the barrier can be 
random,
while lucky cases can be selected later from the record of $I(t)$.

        Direct experiments of this kind as well as experiments on quantum 
feedback control and on Bayesian measurement of entangled qubits (discussed
in sections 7 and 8), are still too difficult for realization at the 
present-day level of technology. In the next two subsections we will 
discuss 
experiments which seem to be realizable (though very hard) at present.

\subsection{Spectral density of the detector current}

        Naively thinking, a qubit with $H\neq 0$ should perform coherent
(Rabi) oscillations with frequency $\Omega$ and these oscillations should
lead to an oscillating contribution of the detector current $I(t)$. 
On the other hand, conventional Eqs.\ (\ref{conv1})--(\ref{conv2}) seem
naively to imply that the qubit eventually reaches a stationary state and 
no oscillations should be present in $I(t)$ after a sufficiently long 
observation. So, it is interesting to find what is the actual spectral 
density of the detector current $S_I(\omega )$ [it is easier to measure 
this quantity experimentally, than to record $I(t)$]. 

        The Bayesian formalism predicts \cite{Kor-osc,Goan-2,Rus-osc} 
the presence of the spectral peak at the Rabi frequency $\Omega$,
however, the height of this peak cannot be larger than 4 times the noise
pedestal. In particular for a symmetric qubit ($\varepsilon =0$)
        \begin{equation}
\frac{S_I(\omega )}{S_0}= 1 +\frac{4\eta} 
{( \omega /\Omega )^2 +(\omega^2-\Omega^2)^2/\Omega^2\Gamma_d^2} \, .
        \label{S(w)} \end{equation}

        Actually, an experimental confirmation of this formula 
would not be a direct verification of the Bayesian formalism, since
Eq.\ (\ref{S(w)}) can be also obtained by other methods, including 
the master equation method \cite{Kor-Av,Kor-osc,Averin} and the method 
based on the Bloch equations \cite{Rus-osc}.

\subsection{Bell-type experiment}

        Another experiment which also seems to be much easier than the 
direct experiments but can unambiguously test the Bayesian formalism, is 
a Bell-type 
experiment in which one qubit is measured by two detectors \cite{Kor-Bell}.
An idea (Fig.\ 3) is to prepare the qubit in a coherent state 
$(|1\rangle +|2\rangle)/\sqrt{2}$, then to switch on the first detector 
($A$) for a relatively short time $\tau_A$ (so that the measurement is 
only partially completed), and to switch on the second detector ($B$)
 a little later. 
If the first measurement changes the qubit state according to the Bayesian 
formalism, then the second measurement can check this change. An output
from a single run of the measurement are two charges $Q_A$ and $Q_B$ 
passed through two detectors. Performing the experiment many times and
analyzing the correlation between $Q_A$ and $Q_B$, one can recover 
the effect of the first measurement on the qubit state \cite{Kor-Bell} (to 
check the change of the nondiagonal matrix element it is necessary to apply 
a $\pi$/2 pulse right after the first measurement). The main advantage of 
this Bell-type experiment in comparison with the direct Bayesian 
experiments
is that the wide bandwidth for the output signal is not necessary;
instead, it is traded for the wide bandwidth of two input lines (switching
detectors on and off), which is much easier to realize experimentally.

\begin{figure}
\centerline{
\epsfxsize=3.1in 
%\hspace{0.5cm}
\epsfbox{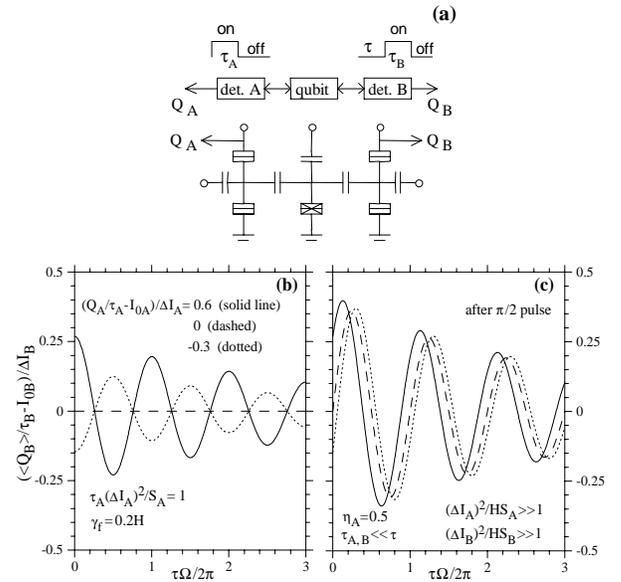}
}  
\vspace{0.3cm} 
\caption{(a) Schematic of the proposed Bell-type correlation experiment
\protect\cite{Kor-Bell}, in which a SCPB qubit is measured by two 
SETs during short time periods $\tau_A$ and $\tau_B$ shifted in time 
by $\tau$. The first measurement leads to an incomplete collapse of 
the qubit initial state $(|1\rangle +|2\rangle )/\sqrt{2}$ 
and affects the result of the second measurement. (b) The average
result $\langle Q_B\rangle$ of the second measurement for a selected 
result $Q_A$ of the first measurement. The sign and amplitude of Rabi 
oscillations depend on $Q_A$, reflecting the change of the diagonal
matrix elements of the qubit density matrix. (c) same as (b) if 
$\pi /2$ pulse is applied immediately after the first measurement. 
Now the phase of oscillations depends on $Q_A$. 
The full-swing oscillations (with amplitude of 0.5 in the ideal case)
indicate a pure qubit state after the first measurement. 
        }
\label{Bell} \end{figure}

\section{Quantum feedback control of a qubit} 

        The Bayesian formalism can be used as a basis for the design 
and analysis of a quantum feedback control of a solid-state qubit. 
As an example,
such feedback control can maintain for arbitrary long time the desired 
phase of a qubit Rabi oscillations, synchronizing them with
a classical reference oscillator, even in presence of 
dephasing environment 
\cite{Kor-rev,Rus-fb}. The overall idea is very close to a
classical feedback loop [Fig.\ 4(a)]. The oscillating qubit evolution is 
monitored by a weakly coupled detector (${\cal C}\equiv 
\hbar (\Delta I)^2/S_0H < 1$), 
the phase $\phi (t)$ of actual Rabi oscillations is compared with the 
desired phase $\phi_0(t)$, and the difference signal $\Delta \phi$ 
is used to control the qubit barrier height. 
If qubit is slightly behind the desired phase, then $H$ is decreased, 
so the oscillations run faster to catch up; if the qubit is ahead of proper
phase, $H$ is increased. It is natural to use a linear control:
$H_{fb}=H(1-F\times\Delta \phi)$, where $F$ is a dimensionless feedback
factor. 

\begin{figure}
\centerline{
\epsfxsize=2.5in 
%\hspace{0.5cm}
\epsfbox{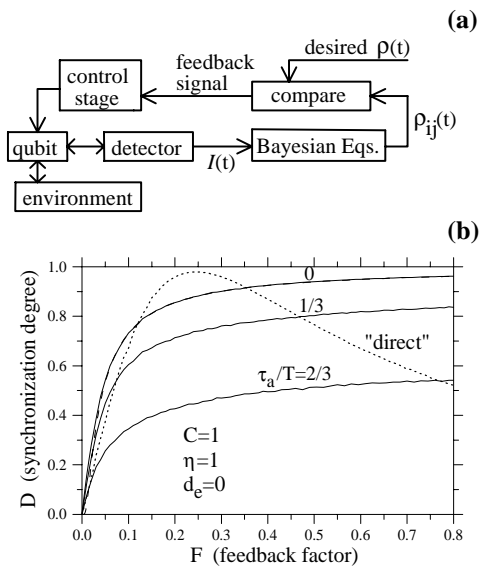}
}  
\vspace{0.3cm} 
%\vspace{4cm}
\caption{ (a) Schematic of the quantum feedback loop maintaining the Rabi
oscillations of a qubit by synchronizing them with a classical harmonic
signal. (b) Solid lines: the synchronization degree $D$ as a function of 
the feedback
factor $F$ for several values of available bandwidth $\tau_a^{-1}$. 
While synchronization can approach 100\% for wide bandwidth, it worsens
when $\tau_a$ becomes comparable to the oscillation period 
$T=2\pi /\Omega$. Dashed line (alsmost coinciding with the upper 
solid line): analytical result
$D=\exp (-{\cal C}/32F)$. Dotted line: synchronization degree for a direct
feedback with $\tau_a=T/10$. 
(From Ref.\ \protect\cite{Rus-fb}.)
        }
\label{feedback} \end{figure}

        The only difference of this loop from a classical feedback is that
even weakly coupled detector disturbs the qubit oscillations. 
However, the induced fluctuations of the oscillation phase are slow, 
and the 
information obtained from the detector happens to be enough to monitor 
the phase fluctuations and compensate them. The quantitative analysis
\cite{Rus-fb} shows that in a limit of good synchronization and absence
of extra environment the qubit correlation function 
$K_z(\tau )\equiv \langle z(t) z(t+\tau )\rangle$ 
(here $z\equiv \rho_{11}-\rho_{22}$) is given by
        \begin{equation} 
K_z(\tau )=\frac{\cos \Omega \tau}{2} \,  \exp \left[
\frac{{\cal C}}{16F} \,
\left( e^{-2FH\tau /\hbar} -1\right) \right] , 
        \label{K_z}\end{equation} 
and does not decay to zero at $\tau \rightarrow \infty$. Correspondingly,
the degree of the qubit synchronization, 
$D\equiv 2\langle \rho \rho_d\rangle -1$ (here $\rho_d$ is the desired
density matrix corresponding to ideal oscillations) is found to be
$D=\exp (-{\cal C}/32F)$ and approaches 100\% at $F \gg {\cal C}$.

        The quality of the qubit oscillations synchronization decreases 
with the decrease of available feedback bandwidth $\tau_a^{-1}$ 
[Fig.\ 4(b)].
It also decreases when the qubit is dephased by an extra environment. 
 For a weak dephasing rate $\gamma_e$ we found numerically \cite{Rus-fb}
a dependence   $D_{max}\simeq 1-0.5 d_e$ where 
$d_e\equiv \gamma_e/[(\Delta I)^2/4S_0]$. This means that the feedback
loop can efficiently suppress the qubit dephasing due to coupling to the 
environment if this coupling is much weaker than the coupling to a nearly
ideal detector. 

        Besides the linear feedback $H_{fb}=H(1-F\times\Delta \phi)$,
we have also studied the ``direct'' feedback 
$H_{fb}(t)/H-1=F\{2[I(t)-I_0]/\Delta I -  \cos \Omega t\} \sin \Omega t$ 
and found that it can also provide a good phase synchronization if
$F/{\cal C}$ is close to 1/4 [Fig.\ 4(b)]. 
The direct feedback is much easier for 
an experimental realization because it does not require  solving 
the Bayesian equations (\ref{Bayes1m})--(\ref{Bayes2m}) in real time.

\section{Bayesian measurement of entangled qubits}

        The Bayesian formalism has been generalized to a continuous quantum
measurement of entangled qubits in Ref.\ \cite{Kor-ent}. Suppose
a detector is coupled to $N$ entangled qubits. In the ``measurement'' basis
there are $2^N$ states $|i\rangle$ of the qubits which correspond to up to 
$2^N$ different dc current levels $I_i$ of the detector (some of these
currents can coincide, for example, if two or more qubits are coupled 
equally strong to the detector). It has been shown that the generalization 
of Eqs.\ (\ref{Bayes1m})--(\ref{Bayes2m}) for this case is 
\cite{Kor-ent} 
        \begin{eqnarray}
&& \dot{\rho}_{ij} = \frac{-\imat}{\hbar} [H_{qb},\rho ]_{ij} + 
\rho_{ij} \, \frac{1}{S_0} \sum_k \rho_{kk} \left[ 
\left(I(t)-\frac{I_k+I_i}{2}\right) 
\right. 
        \nonumber \\
&& \hspace{0.3cm}  \left. 
\times (I_i-I_k) 
+ \left(I(t)-\frac{I_k+I_j}{2}\right) (I_j-I_k) \right] 
-\gamma_{ij}\rho_{ij} ,
        \label{gen-gen}\end{eqnarray}
where the first term describes the unitary evolution due to the 
Hamiltonian of qubits $H_{qb}$ and 
        \begin{equation}
\gamma_{ij}=(\eta^{-1}-1)(I_i-I_j)^2/4S_0, 
        \end{equation}
while Eq.\ (\ref{I(t)}) is replaced by 
        \begin{equation}
I(t) =\sum_i \rho_{ii} (t) I_i +\xi (t).
        \end{equation}
Notice that there is no mutual decoherence ($\gamma_{ij}=0$) between 
states $|i\rangle$ and $|j\rangle$ even for a nonideal detector if the 
corresponding classical currents coincide, $I_i=I_j$. 
This is because  the detector noise cannot destroy the coherence
between states which are equally coupled to the detector.

        Translating Eq.\ (\ref{gen-gen}) from Stratonovich form into 
It\^o form, we get 
        \begin{eqnarray}
&& \dot{\rho}_{ij} = \frac{-\imat}{\hbar} [H_{qb},\rho ]_{ij} + 
\rho_{ij} \, \frac{1}{S_0} 
\Big( I(t)-\sum_k\rho_{kk}I_k \Big) 
        \nonumber \\
&& \hspace{0.5cm}
\times \Big( I_i+I_j-2\sum_k\rho_{kk}I_k \Big) 
 -\Big( \gamma_{ij}+\frac{(I_i-I_j)^2}{4S_0}\Big)  
\rho_{ij} , 
        \label{gen-Ito}\end{eqnarray}
while in the ensemble-averaged equations the second term of Eq. 
(\ref{gen-Ito}) (which depends on $I(t)$) is averaged to zero.

        These Bayesian equations have been applied in Ref.\ \cite{Rus-2qb} 
to the analysis of a simple setup (Fig.\ 5) in which a detector is
equally coupled to two similar qubits (both qubits are symmetric,
$\varepsilon_a=\varepsilon_b=0$, and do not interact directly with each 
other). An interesting effect has been found in the case when the Rabi
frequencies $\Omega_a=2H_a/\hbar$ and $\Omega_b=2H_b/\hbar$ exactly 
coincide.
Then there are two possible scenarios of the two-qubit evolution due to 
measurement, starting from a general mixed state. Either qubits become  
fully entangled collapsing into the Bell state 
$(\mid \uparrow_a\downarrow_b\rangle -\mid \downarrow_a\uparrow_b\rangle )
/\sqrt{2}$
(we call this process spontaneous entanglement),
or the state falls into the orthogonal subspace of the two-qubit
Hilbert space. Experimentally these two scenarios can be distinguished 
by different spectral density $S_I(\omega )$ of the detector current
[Fig.\ 5(c)].
In the case of Bell state, $S_I(\omega )$ is just the flat noise $S_0$ of 
the detector because the signals from two qubits compensate each other, 
while in the other scenario $S_I(\omega )$ has a spectral 
peak at the Rabi frequency, which height is equal to $32\eta /3$ 
for a weakly coupled detector (${\cal C}_a={\cal C}_b < 1$) \cite{Rus-2qb}. 

\begin{figure}[t]
\centerline{
\epsfxsize=2.8in 
%\hspace{0.5cm}
\epsfbox{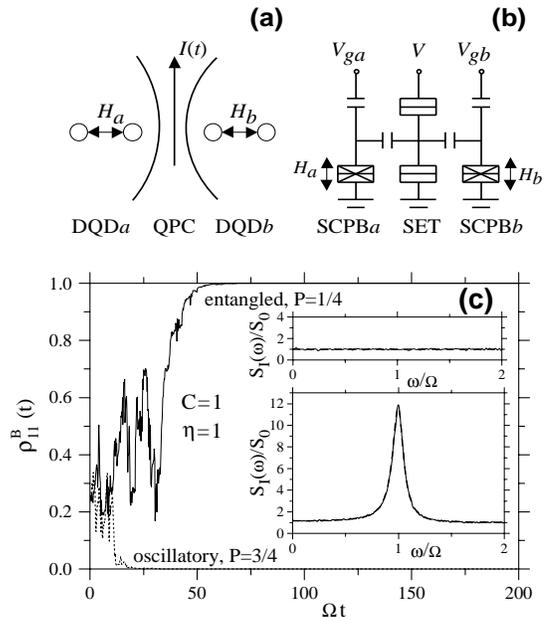}
}  
\vspace{0.3cm} 
\caption{(a) Two qubits made of double quantum dots measured by an 
equally coupled quantum point contact. (b) Similar setup made of
single-Cooper-pair boxes measured by a single-electron transistor.
(c) Two Monte Carlo realizations of the two-qubit state evolution starting 
from the fully mixed state for a symmetric setup ($\rho_{11}^B$ is 
the diagonal component of the two-qubit density matrix, corresponding to 
the Bell state 
$(\mid \uparrow_a\downarrow_b\rangle -\mid \downarrow_a\uparrow_b\rangle )
/\sqrt{2}$). With probability 1/4 
the qubits become fully entangled, $\rho_{11}^B\rightarrow 1$ 
(``spontaneous
entanglement''); then the detector output is a pure noise (upper inset). 
With probability 3/4 the state is gradually collapsed into the orthogonal
subspace, $\rho_{11}^B\rightarrow 0$; then the detector signal shows a 
spectral peak at the Rabi frequency $\Omega$ with the peak-to-pedestal
ratio of 32/3. 
(From Ref.\ \protect\cite{Rus-2qb}.) 
        }
\label{entanglement} \end{figure}

        The probabilities of two scenarios are equal to the contributions
of two subspaces in the initial state $\rho (0)$; for the case of fully
mixed initial state they are equal to 1/4 and 3/4, respectively. 
The considered
setup can obviously be used for a {\it preparation of the Bell state}
without direct interaction between two qubits. Notice that if the state
collapsed into the orthogonal subspace, we can apply some noise which
affects $\varepsilon_a$ and/or $\varepsilon_b$ and therefore mixes
the two-qubit density matrix, and try the measurement again. In this way
the probability $1-(3/4)^M$ to obtain the Bell state can be made arbitrary
close to 100\% by allowing sufficiently large number $M$ of attempts. 

     In actual experiment the symmetry of the setup cannot be made exact.
In this case the Bell state and the oscillating state are not infinitely
stable and there will be switching between them. The calculations show 
\cite{Rus-2qb} that the switching rate $\Gamma_{B\rightarrow O}$ from
the Bell state into the oscillating state is equal to
$\Gamma_{B\rightarrow O} = (\Delta\Omega)^2/2\Gamma_d$ due to
slightly different Rabi frequencies 
[$\Gamma_d =\eta^{-1}(\Delta I)^2/4S_0$],
$\Gamma_{B\rightarrow O} = (\Delta {\cal C}/{\cal C})^2\Gamma_d /8$ due to
slightly different coupling, and 
$\Gamma_{B\rightarrow O} = (\gamma_a+\gamma_b)/2$ 
due to an extra environment acting on two qubits separately. The rate 
of the return switching is 3 times
smaller, $\Gamma_{O\rightarrow B}=\Gamma_{B\rightarrow O}/3$. 
Notice that in this case the averaged height of the Rabi spectral peak 
is equal to $8\eta S_0$, which is exactly twice as much as for 
a single qubit. 

        Even though such experiment on spontaneous entanglement is still 
extremely difficult for a realization, it should be noted that for the
observation of the phenomenon the detector quantum efficiency $\eta$ 
should not necessarily be close to 100\%; it should only be large
enough to allow distinguishing the Rabi spectral peak with the
peak-to-pedestal ratio of $32\eta /3$.

\section{Discussion} 

        In this paper we have reviewed the basic derivation and 
some applications of the Bayesian approach to continuous quantum 
measurement of solid-state qubits. Even though this is a new 
subject for the solid-state community, many similar formalisms have
been developed in other fields of quantum physics. Generally, this 
type of approach which takes into account the measurement outcome, 
is usually called selective or conditional quantum measurement. 
However, there is a rather broad variety of formalisms and their  
interpretations within the approach (for example, see reviews 
\cite{Carmichael,Plenio,Mensky,Presilla}). Some of key words related to 
this subject are: quantum trajectories, quantum state diffusion, quantum
jumps, weak measurements, stochastic evolution of the wavefunction, 
stochastic Schr\"odinger 
equation, complex Hamiltonian, restricted path integral, quantum
Bayes theorem, etc. The approach of conditional quantum measurements
is relatively well developed in quantum optics. In particular, 
the optical quantum feedback control has been well studied theoretically 
(see, e.g.\ \cite{Caves,Wiseman-fb,Tombesi,Doherty,Wiseman-Bfb}) 
and was recently 
realized experimentally \cite{Armen}. In relation to continuous quantum 
measurement of single systems, the quantum optics seems to be about 
10 years ahead of the solid-state physics. However, the interest 
to this problem in the solid-state community has significantly
increased after the ``which path'' experiments \cite{Buks,Sprinzak}. 
Quite possibly it will be a rapidly growing field in the nearest future,
especially because of its direct relation to the solid-state quantum
computing.

        In this paper we have discussed two solid-state experiments 
which seem to be realizable (though very difficult) today. First,
it would be interesting to measure the spectral density of the detector
current when the measured qubit performs coherent oscillations,  
and compare experimental results with the theoretical prediction
that the spectral peak in the best case is 4 times higher than the noise
pedestal. Second, the Bell-type correlation experiment with one qubit 
measured by two detectors would be able to verify that the qubit state
remains pure during the whole measurement process and show the possibility
of monitoring the qubit evolution precisely. This would be the first step
towards realization of the quantum feedback control of solid-state qubits.
A continuous monitoring of entangled qubits would be another 
very interesting experiment. Hopefully, the rapid progress of solid-state 
technology will make these experiments possible in a reasonably near
future.

The author thanks  D.\ Averin, E.\ Buks, G.\ Milburn, and R.\ Ruskov  
for useful discussions. 
The work was supported by NSA and ARDA under ARO grant DAAD19-01-1-0491.

\end{document}